\documentclass[sigplan, 10pt, nonacm]{acmart}

\usepackage{amsmath}
\usepackage{amsthm}
\usepackage{siunitx}

\usepackage{caption}
\usepackage{subcaption}

\usepackage{pgfplots}
\usepackage{pgfmath}
\pgfplotsset{compat=1.18}

\usepackage{url}
\usepackage{listings, lstautogobble}
\usepackage{xspace}
\usepackage{ifthen}

\usepackage{graphicx}
\usepackage{booktabs}

\usepackage{hyperref}
\usepackage[nameinlink]{cleveref}

\newcommand{\pando}{\textsc{Pando}\xspace}
\newcommand{\bosque}{\textsc{Bosque}\xspace}

\newcommand{\wscheat}{\vspace{-1mm}}

\newboolean{reviewversion}
\setboolean{reviewversion}{true} 

\newcommand{\rvc}[2]{\ifthenelse{\boolean{reviewversion}}{#1}{#2}}
\ifthenelse{\boolean{reviewversion}}{}{}

\newcommand{\pnintyninetime}{\us{300}}

\newcommand{\eg}{\hbox{\emph{e.g.}}\xspace}

\newcommand{\etc}{\hbox{\emph{etc.}}\xspace}
\newcommand{\vs}{\hbox{\emph{vs.}}\xspace}

\newcommand{\kb}[1]{\hbox{\qty{#1}{\kilo\byte}}\xspace}
\newcommand{\mb}[1]{\hbox{\qty{#1}{\mega\byte}}\xspace}
\newcommand{\gb}[1]{\hbox{\qty{#1}{\giga\byte}}\xspace}

\newcommand{\us}[1]{\hbox{\qty{#1}{\micro\second}}\xspace}

\newboolean{showcomments}
\setboolean{showcomments}{true} 
\ifthenelse{\boolean{showcomments}}{
  \newcommand{\nbc}[3]{
    {\colorbox{#3}{\bfseries\sffamily\scriptsize\textcolor{white}{#1}}}%
    {\textcolor{#3}{\textsf\small$\blacktriangleright$\textit{#2}$\blacktriangleleft$}}}
  \newcommand{\todo}[1]{\nbc{TODO}{#1}{red}\xspace}
}{
  \newcommand{\nbc}[3]{}
  \renewcommand{\todo}[1]{}
}

\newcommand{\cf}[1]{\texttt{#1}}

\usepackage{color}
\definecolor{purple}{RGB}{75, 0, 255}
\definecolor{cgreen}{rgb}{0.25,0.5,0.35} 

\lstdefinelanguage{bosque}{
keywords={concept, entity, datatype, action, api, typedecl, enum, type, provides, field, env, switch, match, abstract, method, if, then, elif, else, function, yield, return, true, false, none, let, var, in, requires, ensures, invariant, validate, recursive, sensitive, using, of, this, pred, fn, ref, examples, for, defer, test, const, override},
keywordstyle=\color{blue}\bfseries,
identifierstyle=\color{black},
alsoother={@},
sensitive=true,
comment=[l]{\%\%},
commentstyle=\color{cgreen}\bfseries
}
 
\acmConference[LCTES '26]{The 27th ACM SIGPLAN/SIGBED International Conference on Languages, Compilers, Tools and Theory of Embedded Systems}{June 2026}{Boulder, Colorado, United States}
\acmYear{2026}
\acmISBN{978-x-xxxx-xxxx-x/26/06}
\acmDOI{10.1145/xxxxxxx.xxxxxxx}
\copyrightyear{2026}
\setcopyright{none}

\begin{document}

\title{Embedded Made Easy -- Rethinking Embedded + Cloud Software Development (WIP)}

\author{Anthony Arnold}
\email{anthony.arnold@uky.edu}
\affiliation{%
  \institution{University of Kentucky}
  \city{Lexington}
  \state{Kentucky}
  \country{USA}
}

\author{Mark Marron}
\email{mark.marron@uky.edu}
\affiliation{%
  \institution{University of Kentucky}
  \city{Lexington}
  \state{Kentucky}
  \country{USA}
}

\begin{abstract}
The process of engineering and deploying applications in the edge/embedded space is massively complicated by the non-homogeneous nature of the software stack and the complexity of 
diagnostics \& debugging. Often different languages and runtimes are used for different components of the system forcing designers to, irrevocably, make decisions about what 
components run on the periphery and what components run in the cloud. Further complications arise when handling and diagnosing failures in the system. Multiple 
stacks and, often, limited support for debugging complicate the already difficult task of analyzing distributed applications.

This paper presents a work-in-progress vision for a unified language and runtime system that allows applications to scale seamlessly across the edge and cloud. 
Using a single language and runtime, applications can be developed and tested in a single environment, and then deployed to any component of the system 
-- from resource limited controllers to large cloud servers. Further, we outline how this retargetable stack can provide integrated diagnostics and 
debugging tools that allow developers to record and replay distributed events locally for analysis and debugging.
\end{abstract}

\maketitle

\section{Introduction}
\label{sec:intro}

The process of building, deploying, and maintaining applications across the edge-to-cloud continuum is traditionally plagued by fragmented software stacks. 
Developers are forced to use different programming languages, runtimes, and ecosystems depending on whether a component is targeted for a resource-constrained peripheral device 
or a high-performance server. This fragmentation not only forces premature and irrevocable architectural decisions but also severely complicates the already daunting tasks 
of handling errors and diagnosing failures in distributed systems.

This paper presents a work-in-progress vision for a unified language and runtime system designed to eliminate these boundaries and seamlessly scale software from the smallest 
edge devices to the largest cloud nodes. By leveraging a single language and runtime environment, developers can author and test applications uniformly before deploying the exact 
same code to any hardware component within the system architecture.

This paper outlines our overarching vision and details the following key contributions:
\begin{itemize}
  \item We introduce a unified runtime capable of seamlessly scaling application resources up or down. Utilizing a novel garbage collection architecture, the system efficiently manages 
  memory footprints ranging from a few hundred kilobytes on embedded controllers to gigabytes on cloud servers, while concurrently scaling execution from single-threaded asynchronous 
  tasks to large multi-threaded pools.
  \item We detail a framework that enforces fully deterministic and platform-independent language semantics. Explicitly moderating all external platform interactions and dependencies 
  through a neutral interface layer ensures identical (bit-precise) behavior across vastly different hardware targets.
  \item We outline language and runtime mechanisms dedicated to resilient execution and system diagnostics. This includes specialized control flow constructs for cleanly 
  separating success and failure paths, as well as built-in support for trace recording that enables developers to locally replay and diagnose complex, distributed 
  execution failures.
\end{itemize}

The key to our retargetable stack is the \bosque~\cite{bosque,bsqon,highassurance} programming language and associated \pando~\cite{pando,agentic} runtime. The \bosque language is designed to be a high-level, 
general-purpose programming language. For our purposes, the key features of the language are a fully deterministic core, entirely referentially transparent semantics, 
and all external interactions are explicitly moderated. 

\wscheat
\section{Automatic Scaling of Software}
\label{sec:scaling}
A central challenge for automatic scaling of software across the edge/cloud continuum is dealing with memory footprint and thread counts. The difference in scale from a simple controller, 
perhaps only a \mb{} or two of memory and $1-2$ threads, to a large cloud server with \gb{} of memory and hundreds of threads is enormous. There is, seemingly, a fundamental tension between 
a resource efficient programming language and runtime (like C) that can fit within the limits of the small system, and a high-productivity and scalable programming language and runtime 
(like GC languages) that allow for faster development and take advantage of the plentiful resources of the large system. This is further complicated by 
conflicting latency \vs throughput requirements of the different components of the system -- where embedded code often must be responsive and non-blocking while backend services are focused 
on long-running data processing tasks.

\wscheat
\subsection{Scaling Memory Management}
A recent breakthrough in garbage collection~\cite{pando} provides a path forward for a single language and runtime that can scale across the edge/cloud continuum. Leveraging 
a combination of distinctive features in the \bosque programming language, cycle-free data structures and complete referential transparency, the \pando runtime provides 
a collector that simultaneously provides the following guarantees:
\begin{itemize}
  \item \textbf{Bounded Collector Pauses:} The collector only requires the application to pause for a bounded (small constant) period.
  \item \textbf{Starvation Freedom:} The collector can never be outrun by the application allocation rate and will always satisfy allocation requests (until true exhaustion).
  \item \textbf{Constant Memory Overhead:} The reserve memory required by the allocator/collector is bounded by a (small) constant overhead.
\end{itemize}

These properties allow the \pando runtime to scale the resources required to run a \bosque application up and down as needed. 
The results published in \cite{pando} show that, on a well-resourced compute node, the \pando collector imposes an overhead on the order of $5-10\%$ 
while providing astonishing p99 pauses under \pnintyninetime. When compared to a state-of-the-art Java GC/runtime, using an allocation intensive workload, the single threaded \pando 
collector is approximately half the throughput of the multi-threaded JVM runtime \emph{but} uses less than $1.5\times$ the application live memory as opposed to $8\times$-$12\times$(!) for the 
Java GC. When constraining the GCs to a smaller $2\times$ max footprint (\mb{400}) then the JVM performance drops precipitously and the \pando collector is approximately $2\times$ faster than 
the JVM version!

For scaling down to edge/embedded applications we extended this prior work to run with even smaller nurseries and auxiliary data structures. This 
preliminary work already allows the \pando GC to run effectively in a nursery as small as \kb{256}, enabling us to run with a total memory overhead of 
$O(\kb{384})$, which easily fit in the constraints of any powered board and many larger SoC controllers. With this GC in place, we can now take a single application codebase 
written in \bosque and, without any changes to the code, deploy and execute it on hardware with anywhere from \mb{1} to \gb{100} with guaranteed GC performance and behavior.

\wscheat
\subsection{Scaling Threads}
The \bosque language provides a simple and efficient concurrency model based on lightweight \cf{Tasks}~\cite{agentic} and structured concurrency. As this parallelism model 
is convertible to continuation passing we can transparently map it onto a, compiler and runtime directed, combination of 
lightweight async tasks and thread pool. This allows us to scale the thread count of the application up and down as needed by the underlying hardware as well as picking the model of 
execution best suited for the Task structure -- \eg all IO is async (using io\_uring~\cite{iouring}) while pure compute can be run on a dedicated thread without inturruption. On a small controller we can run with 
a single thread and leverage the async tasks to provide concurrency while on a large server we can use a large thread pool to maximize performance. Again, this is all transparent 
to the application code -- the same code can be deployed to either system without modification.

The current \pando GC implementation supports multi-threaded GC via classic locking and stop-the-world collection. However, we are currently working on a concurrent version of the GC 
that, again, leveraging unique features of the \bosque language, will allow us to eliminate almost all locking and most scenarios that could lead to false-sharing cache eviction conflicts. 
With this concurrent GC in place we will be able to transparently scale the thread count of the application up and down as needed without GC performance degradation or changes to the 
underlying application code as well. This design allows us to take advantage of the plentiful resources of a large server without compromising our ability to run on a small systems.

\wscheat
\section{System Encapsulation}
\label{sec:encapsulation}
Even with a single language and runtime that can scale across the edge/cloud continuum, from the standpoint of the application code and runtime system, there are still challenges in 
ensuring that applications can be moved between different components of the system without extensive effort. Specifically, languages often have platform specific semantics that can 
lead to differences in behavior when code is moved between different hardware and software environments. More generally, software ecosystems are often built with expectations of 
access to various platform features which can make it difficult or infeasible to re-target a deployment.

\wscheat
\subsection{Language Semantic Stability}
Language semantics often expose platform specific details that can lead to differences in application behavior, such as endianness or pointer identity, and even when hidden 
via core language semantics, these often leak out via foreign-function interfaces (FFI) or libraries. \bosque takes a hard-line approach to this problem by ensuring the 
core language semantics are entirely platform independent, in fact \bosque ensures that all (non-Task code) is fully deterministic! 

\bosque also strongly disallows direct FFI and moderates any external calls through a neutral interface layer (BAPI~\cite{bsqon}). This design ensures that that all platform 
specific interactions are explicitly moderated via a well-defined interface. Thus, the same \bosque code can be deployed to different platforms without any changes and 
all executions are guaranteed to have the the same behavior -- up-to Task ordering they will produce the, bit precise, same results on every run on any platform! 

The BAPI interface layer also supports automated generation of standard polyglot RPC bindings. In the most common scenario, RESTful systems, this is JSON encoding 
with a JSON-RPC~\cite{jsonrpc} or OpenAPI~\cite{openapi} specification. This allows us to easily integrate with other languages and systems and to auto-generate all the 
associated, and error prone~\cite{restler}, parsing and data-validation logic.

\wscheat
\subsection{Platform Dependency Encapsulation}
Even with platform independent language semantics, applications often have dependencies on platform specific features, such as file systems, network stacks, \etc. The \bosque 
system takes a strict approach to this, inspired by JavaScript~\cite{chromium,napi} and the success in embedding it in a variety of environments. \bosque defines a core set of platform independent 
libraries that provide a rich-set of common data types and functionality (all independent of platform specific features). Then, using BAPI~\cite{bsqon} all platform specific features 
are exposed via independent packages, including the file system, network, \etc. Application code can then explicitly import and use these as needed, but they are not required for 
the core application logic, and they can be entirely removed if not relevant. This design makes \bosque ideal for a REST first ecosystem where applications are expected to be built 
as a collection of services that interact via well-defined APIs. By ensuring that all platform specific features are explicitly encapsulated and moderated, we avoid the need to 
implement and load complex code on small systems that may not have the resources to support it, \eg full file systems. 

The combination of platform independent language semantics and explicit platform dependency encapsulation enables us to take a single codebase and trivially re-target components 
of the system to different hardware targets. The semantic stability ensures that, regardless of the underlying hardware and software environment, the application code will produce the, 
bit precise, identical results on any platform! The platform dependency encapsulation allows us to aggressively shake out platform specific dependencies and, even in our current 
prototype which depends on C++ \cf{std} more than we would like, the resulting binaries can be kept in the range of a few \kb{100} to a few \mb{} in size. As with the compact RAM footprint this ensures 
that code written in the high-level \bosque language can be deployed to low-resource embedded systems without extensive reworking.

\wscheat
\section{Integrated Fault Handling \& Diagnostics}
\label{sec:diagnostics}
Even the most carefully designed systems will encounter faults and failures in the field. The complexity of distributed applications, especially those that span the edge/cloud continuum and 
involve multiple software stacks, makes diagnosing and handling these failures particularly difficult. Applications in this space must contend with a variety of failure modes ranging from 
software errors to intermittent network issues. Oddly, most programming languages and runtimes provide very little support for handling failures and focus most features on supporting 
``happy path'' execution. When failures inevitably occur, developers are often left with a primitive set of tools, such as logging and core dumps, to try to diagnose the issue 
after the fact. 

\wscheat
\subsection{Integrated Fault Handling and Recovery}
The core \bosque compute language is fully referentially transparent which guarantees that all values are immutable. Thus, failure recovery code can always assume 
data has not been compromised during a failure and that the values are unchanged from before the failing code was run. This allows recovery code to proceed safely and without 
concern for partially modified values or other corrupted state. 

\bosque extends the core compute language with a Task based concurrent programming model that provides \emph{structured parallel} execution. Based on the effectiveness of automated 
error capture and conversion of logic errors in JavaScript Promises~\cite{jspromise}, the \bosque Task model provides a similar mechanism for automatically capturing and converting logic 
errors in concurrent code to structured failure objects that can be handled programmatically. When combined with special control flow constructs for managing failure paths, developers 
can easily construct clean success and failure logic paths instead of convoluted code with intermixed failure checks and application logic. 

\begin{figure}[t]
\centering
\begin{lstlisting}[language=bosque]
action doit(): ByteBuffer {
  let data = Task::run<Fs::ReadFile>("file.txt");
  match(data) {
    APIResult::Success => { return data.value; }
    _ => { yield APIResult::error("Failed read"); }
  }
}

api main(): APIResult<Int, String> {
  let result = doit();
  return APIResult::success(result); 
}
\end{lstlisting}
\vspace{-3mm}
\caption{Using the \cf{yield} construct to manage success and failure paths in \bosque.}
\label{fig:yield}
\vspace{-5mm}
\end{figure}

\Cref{fig:yield} shows an example of how the \bosque \cf{yield} construct can be used to manage success and failure paths in a structured way.
If the file read fails, the \cf{yield} will unwind the call stack and return the given value (\cf{APIResult::error}) as the result of the \cf{main} api call, skipping 
all the remaining code in the \cf{doit} and \cf{main} functions. If the file read succeeds, then the success path is taken and the file data is returned, via the normal control flow, as 
the result of the \cf{doit} and then \cf{main} api call. Thus, the error handling logic and normal application logic can be cleanly separated into distinct parts of the codebase allowing 
for simpler and more maintainable fault management code. 

\wscheat
\subsection{Live Trace Recording}
As faults will inevitably occur in the field, especially in complex distributed applications, it is critical to have tools that allow developers to diagnose and analyze these failures. 
The \bosque language and \pando runtime provide built-in support for recording and replaying execution traces of applications in a completely platform agnostic manner. As all operations 
with the environment are explicitly moderated via the BAPI interface, the runtime can automatically capture all interactions with the environment, including inputs and outputs, and 
efficiently store them in a replay-log~\cite{ambrosia,ttdjs}. By bootstrapping on the GC, the runtime can also capture efficient state snapshots of the applications logical 
(not just physical) memory state at various points during execution. This logical model allows the system to inflate the snapshot into a executable state even on an entirely different 
hardware platform or operating system~\cite{ttdjs}.

Beyond simply replaying the execution of a single node (component) in the system, the \pando runtime can be configured to automatically log activity on each system and propagate 
correlation identifiers~\cite{correlationid} as deployed services interact with each other. This allows developers to replay the execution of the application, or even an entire sequence of 
distributed events, locally on their development machine and observe the exact behavior that led to a field failure! 

\wscheat
\section{Discussion}
Previous work on unified languages for simplifying embedded development has generally focused on taking an existing high-level language and creating a subset that can 
effectively target small systems~\cite{micropython,devicesript,swiftembedded}. These approaches have had success in allowing developers to write code in a single 
language and deploy it to different components of the system but, inevitably, the difference between the full language and the subset leads to fragmentation and the need for developers 
to make decisions about what code can run on the periphery and what code must run in the cloud.

Related work on runtimes includes approaches to creating uniform runtime systems. In some cases this is done through slimmed down linux images~\cite{azuresphere} which provide 
a standardized environment for embedded and cloud environments at a cost of larger footprints. More aggressive approaches have been taken to create unikernels and library 
OSes~\cite{unikernel} that can be used to create custom runtimes for different applications. However, these approaches can, like the language subset approaches, lead to 
fragmentation. Further, they often increase development complexity as standard debugging and operations tools are not designed to work with these custom runtimes and require additional 
effort to integrate.

Hardware (or kernel) offloading is also related to the vision presented in this paper. 
Just as there are features in \bosque that made it possible to scale across the edge/cloud continuum, these features also make it possible to 
do precise analysis for resource bounds analysis and checking safety to enable offloading. Our preliminary work on these topics indicates that \bosque programs 
are can be used as drop in replacements for scenarios like eBPF~\cite{ebpf} and even generalize to code that uses dynamic allocation and iteration structures. 

\wscheat
\section{Onward!}
The unified language and runtime vision presented in this work creates new opportunities for building and deploying applications in the embedded and cloud space. 
Our current work is focused on the integration of the prototype and research implementations for the capabilities described in this WIP paper into a robust and reliable software 
artifact that can be used by industry and the research community. By providing 
a single environment that scales seamlessly, we can rethink how systems are constructed and enable research into advanced topics such as precise 
resource bounds analysis~\cite{raml}, transparent hardware offloading~\cite{ebpf}, and dynamic process migration~\cite{ambrosia}. 


\bibliographystyle{ACM-Reference-Format}
\bibliography{bibfile}

@inproceedings{bosque,
author = {Marron, Mark},
title = {{Toward Programming Languages for Reasoning: Humans, Symbolic Systems, and AI Agents}},
year = {2023},
series = {Onward!}
}

@misc{pando,
      title={{Catalpa: GC for a Low-Variance Software Stack}}, 
      author={Anthony Arnold and Mark Marron},
      year={2025},
      archivePrefix={arXiv},
      url={https://arxiv.org/abs/2509.13429}, 
}

@misc{agentic,
      title={Toward an Agentic Infused Software Ecosystem}, 
      author={Mark Marron},
      year={2026},
      eprint={2602.20979},
      archivePrefix={arXiv},
      url={https://arxiv.org/abs/2602.20979}, 
}

@INPROCEEDINGS{restler,
  author={Vaggelis Atlidakis and Patrice Godefroid and Marina Polishchuk},
  series={ICSE}, 
  title={{RESTler: Stateful REST API Fuzzing}}, 
  year={2019}
}

@inproceedings{ttdjs,
 author = {Barr, Earl T. and Marron, Mark and Maurer, Ed and Moseley, Dan and Seth, Gaurav},
 title = {{Time-travel Debugging for JavaScript/Node.Js}},
 series = {FSE},
 year = {2016}
}

@inproceedings{raml,
author = {Hoffmann, Jan and Das, Ankush and Weng, Shu-Chun},
title = {{Towards Automatic Resource Bound Analysis for OCaml}},
year = {2017},
series = {POPL}
}

@inproceedings{highassurance,
author = {Stephen Goldbaum and Attila Mihaly and Tosha Ellison and Earl T. Barr and Mark Marron},
title = {{High Assurance Software for Financial Regulation and Business Platforms}},
year = {2022},
series = {VMCAI}
}

@inproceedings{bsqon,
author = {Marron, Mark},
title = {{A Programming Language for Data and Configuration!}},
year = {2024},
series = {Onward!}
}

@TechReport{ambrosia,
author = {Goldstein, Jonathan and Abdelhamid, Ahmed and Barnett, Mike and Burckhardt, Sebastian and Chandramouli, Badrish and Gehring, Darren and Lebeck, Niel and Minhas, Umar Farooq and Newton, Ryan and Ghosh Peshawaria, Rahee and Zaccai, Tal and Zhang, Irene},
title = {{A.M.B.R.O.S.I.A: Providing Performant Virtual Resiliency for Distributed Applications}},
organization = {Microsoft},
year = {2018}
}

@inproceedings{azuresphere,
    author = {Nightingale, Ed},
    title = {{A View from Industry: Securing IoT with Azure Sphere}},
    year = {2019},
    series = {HotMobile}
}

@inproceedings{unikernel,
author = {Madhavapeddy, Anil and Mortier, Richard and Rotsos, Charalampos and Scott, David and Singh, Balraj and Gazagnaire, Thomas and Smith, Steven and Hand, Steven and Crowcroft, Jon},
title = {{Unikernels: Library Operating Systems for the Cloud}},
year = {2013},
series = {ASPLOS}
}

@misc{openapi,
    Key = {OpenAPI},
    Note = {\url{https://swagger.io/specification/}},
    Title = {{OpenAPI 3.0 Format}},
    year = {2025}
}

@misc{jsonrpc,
    Key = {JSON-RPC},
    Note = {\url{https://www.jsonrpc.org/specification}},
    Title = {{JSON-RPC Specification}},
    year = {2025}
}

@misc{micropython,
    Key = {MicroPython},
    Note = {\url{https://micropython.org/}},
    Title = {{MicroPython}},
    year = {2025}
}

@misc{devicesript,
    Key = {DeviceScript},
    Note = {\url{https://github.com/microsoft/devicescript}},
    Title = {{DeviceScript}},
    year = {2025}
}

@misc{swiftembedded,
    Key = {Swift},
    Note = {\url{https://docs.swift.org/embedded/documentation/embedded/}},
    Title = {{Embedded Swift}},
    year = {2025}
}

@misc{napi,
    Key = {N-API},
    Note = {\url{https://nodejs.org/api/n-api.html}},
    Title = {{Node-API}},
    year = {2025}
}

@misc{chromium,
    Key = {Chromium},
    Note = {\url{https://www.chromium.org/}},
    Title = {{Chromium Project}},
    year = {2025}
}

@misc{correlationid,
    Key = {Microsoft Engineering Playbook},
    Note = {\url{https://microsoft.github.io/code-with-engineering-playbook/observability/correlation-id/}},
    Title = {{Correlation ID}},
    year = {2025}
}

@misc{jspromise,
    Key = {JavaScript Promises},
    Note = {\url{https://developer.mozilla.org/en-US/docs/Web/JavaScript/Reference/Global_Objects/Promise}},
    Title = {{JavaScript Promises}},
    year = {2025}
}

@MISC{ebpf,
 Key = {Linux Kernel Offload},
 Note = {\url{https://ebpf.io/}},
 Title = {{eBPF Documentation}},
 year = {2019}
}

@misc{iouring,
    Key = {io_uring},
    Author = {Jens Axboe},
    Note = {\url{https://github.com/axboe/liburing/}},
    Title = {{io\_uring}},
    year = {2025}
}

\end{document}